\documentclass[usegraphix]{mn2e}
 
\include{defn}

\usepackage{amsmath}
\usepackage{amssymb}
\usepackage{times}
\usepackage{graphicx}
\usepackage{subfigure}
\usepackage{epsfig}
\usepackage{txfonts}

\def\re{{R_{\rm e}}}

\def\Fa{{F_{\rm asym}}}
\def\Fc{{F_{\rm cusp}}}
\def\ga{{g_{\rm asym}}}
\def\gc{{g_{\rm cusp}}}
\def\mnras{{MNRAS}}
\def\apj{{ApJ}}
\def\apjl{{ApJL}}
\def\aj{{AJ}}

\def\arcsech{{\rm arcsech}}
\def\arcsec{{\rm arcsec}}
\def\erf{{\rm erf}}
\def\rt{{r_{\rm t}}}
\def\Ha{{\mathcal{H}}}
\def\Hac{{\mathcal{H}_{\mathrm{cusp}}}}
\def\Haf{{\mathcal{H}_{\mathrm{asym}}}}

\begin{document}

\pagenumbering{arabic}

\title[A Very Simple Cusped Halo Model]
  {A Very Simple Cusped Halo Model}

\author[Evans \& Williams]
  {N.W. Evans$^1$\thanks{E-mail:nwe@ast.cam.ac.uk},
   A.A. Williams$^1$
 \medskip
 \\$^1$Institute of Astronomy, University of Cambridge, Madingley Road,
       Cambridge, CB3 0HA, UK}

\maketitle

\begin{abstract}
We introduce a very simple model of a dark halo.  It is a close
relative of Hernquist's model, being generated by the same
transformation but this time applied to the logarithmic potential
rather than the point mass. The density is proportiional to
(distance)$^{-1}$ at small radii, whilst the rotation curve is flat at
large radii. Isotropic and radially anisotropic distributions
functions are readily found, and the intrinsic and line of sight
kinematical quantities are available as simple formulae. We also
provide an analytical approximation to the Hamiltonian as a function
of the actions. As an application, we study the kinematic properties
of stellar haloes and tracers in elliptical galaxies. We show that the
radial velocity dispersion of a power-law population in a galaxy with
a flat rotation curve always tends to the constant value. This holds
true irrespective of the anisotropy or the lengthscales of the dark or
luminous matter.  An analogous result holds for the line of sight or
projected velocity dispersion of a power-law surface brightness
profile.  The radial velocity dispersion of Population II stars in the
Milky Way is a strongly declining function of Galactocentric
radius. So, if the rotation curve is flat, we conclude that the
stellar halo density cannot follow a power-law at large radii, but
must decrease more sharply (like an Einasto profile) or be abruptly
truncated at large radii. Both the starcount and kinematic data of the
Milky Way stellar halo are well-represented by an Einasto profile
with index $m\approx 2$ and effective radius $\approx 20$ kpc.
\end{abstract}


\section{INTRODUCTION}

Hernquist (1990) devised a very simple spherical model with potential-
density pair
\begin{eqnarray}
\psi &=& {GM \over r+a},\nonumber\\
\rho &=& {M \over 2\pi}{a\over r(r+a)^3}.
\label{eq:hernquist}
\end{eqnarray}
Here $M$ is the mass of the model and $a$ is a scalelength related to
the half mass radius. Hernquist's model provides a beguiling
combination of simplicity and realism. It has become well-known as a
representation of elliptical galaxies and bulges. It has also become
widely used in $N$ body calculations, as the forces are analytic and
so ease numerical orbit integrations.  We note that the model is
obtained from the familiar Keplerian potential of a point mass by the
straightforward transformation $r \rightarrow r+a$.

Here, we will provide a very simple spherical model of a galaxy with a
flat rotation curve generated by the exact same transformation -- but
this time applied to the logarithmic potential of the singular
isothermal sphere. This gives the potential-density pair
\begin{eqnarray}
\psi &=& -v_0^2 \log (r+a) + {\rm constant},\nonumber\\
\rho &=& {v_0^2 \over 4\pi G}{2a+r\over r(r+a)^2}.
\label{eq:newlog}
\end{eqnarray}
The similarity with Hernquist's original model is evident.
Henceforth, we choose the constant as $v_0^2\log a$ so that the
potential $\psi$ is zero at the centre.

The model has an asymptotically flat rotation curve. But -- like
Hernquist's model -- it has a $1/r$ cusp in the density. This is of
course highly desirable given the findings of Navarro, Frenk \& White
(1996) that galaxy formation via hierarchical merging in a cold dark
matter (CDM) dominated universe gives rise to density cusps of the
form $\rho \propto 1/r$ at small radii. The NFW profile probably does
not describe the haloes of large galaxies like the Milky Way or M31
(e.g., Binney \& Evans 2001) or small dwarf irregulars and spheroidals
(e.g., Oh et al. 2008; Agnello \& Evans 2012). This is thought to be
the result of baryonic processes that may modify the central regions
(e.g., Pontzen \& Governato 2014). Nonetheless, models with $1/r$
density cusps are still invaluable as an approximation to equilibrium
configurations of dark matter in dissipationless simulations.

In Section 2, we describe the properties of our new halo
model~(\ref{eq:newlog}), providing formulae for the kinematic
quantities, distribution functions and actions. Section 3 looks at the
properties of tracer populations with an eye on the kinematics of
stellar haloes and elliptical galaxies. Section 4 sums up, comparing
our halo model with a number of familiar faces from the literature.

\begin{figure}
    \centering
    \includegraphics[width=2.5in]{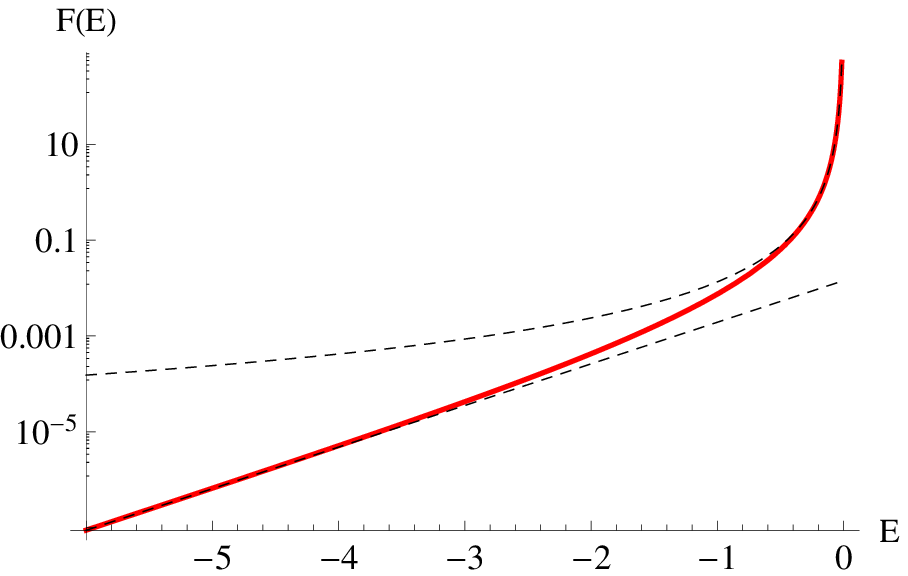}
    \includegraphics[width=2.5in]{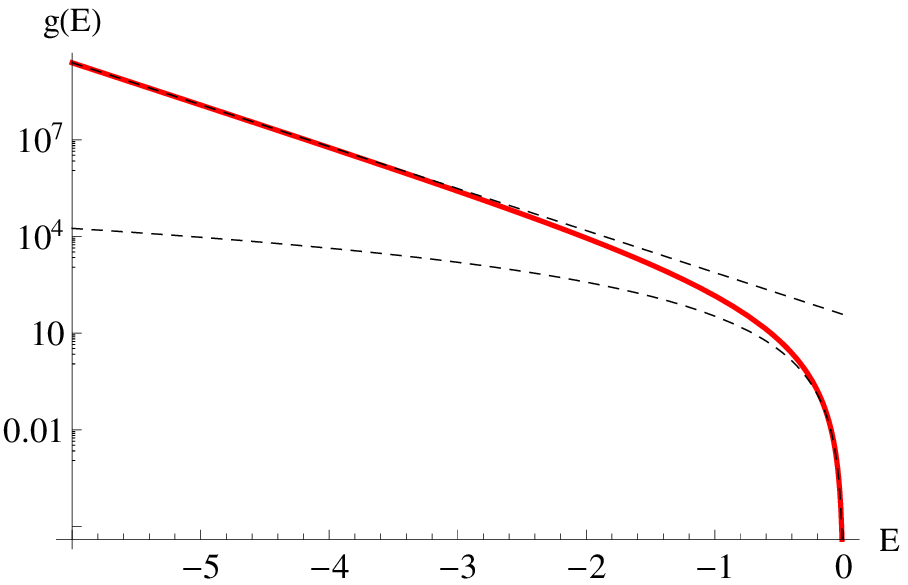}
    \caption{Distribution function $f$ and density of states $g$ as a
      function of binding energy $E$ for the isotropic model
      ($v_0=1=a=G$). The full results are shown in red, whilst the
      asymptotic and cusp approximations are shown as dashed lines.}
    \label{fig:isodf}
\end{figure}
\begin{figure}
    \centering
    \includegraphics[width=3in]{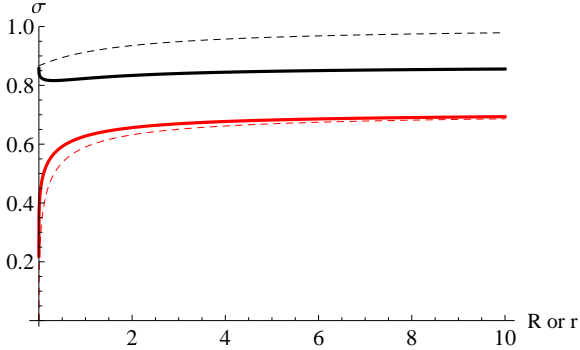}
    \caption{Radial (dashed) and line of sight (unbroken) velocity
      dispersion as a function of intrinsic radius $r$ and projected
      radius $R$ respectively. The isotropic model ($\beta=0$) is
      shown in red and the radially anisotropic model ($\beta = 1/2$)
      in black. Note that the line of sight velocity distribution
      picks up contributions from the tangential velocity dispersion,
      and so the full line lies above the dashed for the isotropic
      model, but below the dashed for the radially anisotropic model.
      Units are chosen so that $v_0=a=1$.}
    \label{fig:vel}
\end{figure}

\section{Properties of the Model}

First, we will derive some properties of the self-consistent
model. Our results apply to the total (baryonic plus dark matter)
content of the galaxy.

\subsection{Preliminaries}

The phase space distribution function (henceforth DF) is of
fundamental dynamical importance. By Jeans (1919) theorem, the DF is a
function of the isolating integrals of motion only. As the model is
spherical, the DF may depend on the binding energy per unit mass $E$
and the modulus of the angular momentum $L$
\begin{eqnarray}
E &=& -\tfrac{1}{2}(v_r^2 + v_\theta^2 + v_\phi^2) - v_0^2 \log (
1+r/a),\nonumber \\
L^2 &=& r^2(v_\theta^2 + v_\phi^2).
\label{eq:angmom}
\end{eqnarray}
Isotropic DFs depend only on $E$, whereas anisotropic DFs depend on $L$
as well. 

We look for DFs in which the anisotropy parameter
\begin{equation}
\beta = 1 - {\langle v_\theta^2 \rangle \over \langle v_r^2 \rangle},
\end{equation}
takes a constant value. Of course, for a spherical density
distribution in a spherical potential, it must be the case that
$\langle v_\phi^2\rangle = \langle v_\theta^2\rangle$. The parameter
$\beta$ may take values in the range $-\infty \le \beta \le 1$. The
isotropic model corresponds to the case $\beta=0$. Models with $\beta
>0$ are radially anisotropic with $\beta \rightarrow 1$ being the
radial orbit model. Models with $\beta <0$ are tangentially
anisotropic with $\beta \rightarrow -\infty$ being the circular orbit
model.

The DF of a spherical system with constant anisotropy is
\begin{equation}
f(E,L)=L^{-2\beta}f_E(E).
\label{eq:ansatz}
\end{equation}
The unknown function $f_E(E)$ can be recovered from the integral
inversion formula (e.g., Wilkinson \& Evans 1999, Evans \& An 2006)
\begin{equation}
f_E(E)=
\frac{2^\beta(2\pi)^{-3/2}}{\Gamma(1-\lambda)\Gamma(1-\beta)}\
\frac{d}{dE}\!\int_0^E\frac{d\psi}{(E-\psi)^\lambda}
\frac{d^nh}{d\psi^n}
\label{eq:disint}
\end{equation}
where $h=r^{2\beta}\rho$ is expressed as a function of $\psi$, and
$n=\lfloor(3/2-\beta)\rfloor$ and $\lambda=3/2-\beta-n$ are the
integer floor and the fractional part of $3/2-\beta$. This includes
Eddington's (1916) formula for the isotropic DF as a special case
($\beta=0$).

If $\beta$ is a half-integer constant (i.e., $\beta=$ $1/2$, $-1/2$,
and so on), the expression for DF further reduces to
\begin{equation}
f(E,L)=\frac1{2\pi^2}\frac{L^{-2\beta}}{(-2\beta)!!}\
\left.\frac{d^{3/2-\beta}h}{d\psi^{3/2-\beta}}\right|_{\psi=E}.
\label{eq:dishalf}
\end{equation}
Rather remarkably, this involves only differentiations. The expression
for the differential energy distribution is
\begin{equation}
\frac{dM}{dE}=f_E(E)g(E)\,
\end{equation}
where the density of states $g(E)$ is
\begin{equation}
g(E) = \frac{(2\pi)^{5/2}\Gamma(1-\beta)}{2^{\beta-1}\Gamma(3/2-\beta)}
\int_0^{r_E}(\psi-E)^{1/2-\beta}r^{2(1-\beta)}dr,
\label{eq:ded}
\end{equation}
generalizing the formula in Binney \& Tremaine (2008) for the
isotropic case. Here, $r_E$ is the radius of the largest orbit of a
particle with energy $E$, that is to say, $\psi(r_E)=E$.

\subsection{Distribution Functions}

The key to the simplicity of our model is that $\psi(r)$ can be easily
inverted to give $r(\psi)$, namely
\begin{equation}
r = a\exp(-\psi/v_0^2) - a.
\end{equation}
This means that $\rho(\psi)$ can also be easily constructed.
\begin{equation}
\rho(\psi) = {v_0^2\over 4 \pi G a^2} { \exp(2\psi/v_0^2)
  + \exp(3\psi/v_0^2) \over 1 -\exp(\psi/v_0^2) }.
\end{equation}
The general constant anisotropy DF can be found using
eq.~(\ref{eq:disint}) as
\begin{equation}
F(E,L) = A L^{-2\beta} \exp((2-2\beta)E/v_0^2) \left[ 1 +
  \sum_{j=1}^{\infty}C_j \exp(jE/v_0^2)\right],
\label{eq:dffull}
\end{equation}
where the constants $A$ and $C_j$ are defined in the Appendix. Note
that the DF is only positive definite if $\beta \le 1/2$. If $\beta$
exceeds this value, the DF is negative at the centre.  This is in
accord with the {\it cusp-anisotropy theorem} of An \& Evans (2005),
which states that in spherical symmetry models with a cusp like
$r^{-\gamma}$ at the centre must satisfy $\beta \le 2\gamma$. 

The asymptotic forms of the DF in the cusp ($r\ll a$) and in the outer
parts ($r\gg a$) are a power-law in energy or an exponential in energy
respectively:
\begin{eqnarray}
\Fc(E,L) &=&
   {v_0^{4-4\beta}\Gamma(5/2-3\beta)\over 2^{5/2-\beta}G a^{2-2\beta}
     L^{2\beta}\Gamma(1-\beta)\Gamma(2-2\beta)} (-E)^{-5/2 +
     3\beta}\nonumber\\
\Fa(E,L) &=& A L^{2\beta} \exp (2-2\beta)E/v_0^2.
\label{eq:DFlim}
\end{eqnarray}

\subsection{Special Cases: Isotropy and Radial Anisotropy}

When $\beta =0$, the isotropic DF is an infinite sum of exponentials
\begin{equation}
F(E) = {\exp(2E/v_0^2)\over 4\pi^{5/2}G v_0 a^2} \left[ 1 +
  2^{-1/2}\sum_{j=1}^{\infty}(2 + j)^{3/2}\exp(jE/v_0^2)\right].
\end{equation}
The density of states is
\begin{eqnarray}
g(E) &=& 8 a^3v_0\pi^{5/2}\Biggl[ \sqrt{2} Y(E/v_0^2) -Y (2E/v_0^2) 
\nonumber\\
&+& {\sqrt{2}\over 3\sqrt{3}}Y(3E/v_0^2) 
- {2\over 3} \sqrt{{-2E\over \pi v_0^2}} \Biggr],
\end{eqnarray}
where $Y(s) = \exp(-s)\erf(\sqrt{-s})$. Again, we can develop simple
approximations valid for the far-field and the cusp, namely
\begin{eqnarray}
\gc(E) &=& {256\sqrt{2}\over 105}{\pi^2a^3\over v_0^6}(-E)^{7/2},\\
\ga(E) &=& {8\sqrt{2}\over 3\sqrt{3}} \pi^{5/2} v_0a^3\exp(-3E/v_0^2).
\end{eqnarray}
Plots of the isotropic distribution function and density of states are
shown in Fig.~\ref{fig:isodf}, together with the approximations that
hold good in the cusp and in the far-field.

The isotropic ($\beta=0$) velocity dispersions are
\begin{equation}
\langle v_r^2\rangle = \langle v_\theta^2\rangle = \langle v_\phi^2\rangle = 
{v_0^2\over 2}\, {4r(a+r)^2 \log \left( 1+ r/a \right)
  -ar(5a+4r) \over a^2(2a+r)}.
\end{equation}
As $r \rightarrow 0$, the velocity dispersions tend to zero, whilst as
$r \rightarrow \infty$, the velocity dispersions tend to
$v_0/\sqrt{2}$.

When $\beta =1/2$, the DF is particularly simple, as the infinite
series reduces to just two terms. (In fact, whenever $\beta$ is a
half-integer, the DF has a simple closed form). We find that
\begin{equation}
F(E,L) = {1 \over 8 \pi^3 G a |L|} \left[ \exp ( E/v_0^2) +
  2\exp(2E/v_0^2)\right].
\end{equation}
This is analogous to the very simple DF for the Hernquist model (Baes
\& Dejonghe 2002, Evans \& An 2005). Evidence from numerical
simulations suggests that the velocity distribution of the dark matter
is radially anisotropic with $\beta \approx 1/2$ (e.g., Hansen \&
Moore 2006), so this is a cosmologically realistic DF. The density of
states is
\begin{equation}
g(E) = 4\pi^3 a^2 (\exp(-E/v_0^2) -1)^2.
\end{equation}
For the radially anisotropic ($\beta = 1/2$) model, the velocity
dispersions are
\begin{equation}
\langle v_r^2 \rangle = {v_0^2\over 2}{3a+2r\over 2a+r },\qquad
\langle v_\theta^2 \rangle = \langle v_\phi^2\rangle = {v_0^2\over 4}{3a+2r\over 2a+r }.
\end{equation}
As $r\rightarrow 0$, the radial velocity dispersion tends to
$\sqrt{3}v_0/2$, whilst as $r \rightarrow\infty$, it tends to $v_0$.

Let us check that our models obey the virial theorem. In both cases,
the potential energy is
\begin{eqnarray}
\Omega (r)&=& 4\pi \int_0^r r^3 \rho {d\psi \over dr} dr  \\
          &=& -{v_0^4\over 2G (a+r)^2}\left[r(2a^2+3ar+2r^2) -2a(a+r)^2 \log (1+r/a) \right]\nonumber.
\end{eqnarray}
The kinetic energy of the isotropic ($\beta =0$) model is
\begin{eqnarray}
T(r) &=& 6\pi \int_0^r r^2\rho\langle v_r^2 \rangle dr
     = {v_0^4\over 4 G a^2 (a+r)}\left[ ar(2a^2+ar-4r^2)\right. \nonumber \\
  &+& 4(a+r)r^3\log( 1+a/r)
   - \left. 2(a+r)a^3\log (1 +r/a)\right],
\end{eqnarray}
so that, as $r\rightarrow \infty$, $T\rightarrow 3v_0^4r/(4G)$ and
$\Omega \rightarrow -v_0^4r/G$.  

By contrast, the kinetic energy of the radially anisotropic ($\beta
=1/2$) model is
\begin{eqnarray}
T(r) &=& (6-4\beta)\pi \int_0^r r^2\rho\langle v_r^2 \rangle dr \nonumber\\
     &=& {v_0^4\over 2G (a+r)}\left[r(a+2r) -a (a+r)\log (1 +r/a)\right]
\end{eqnarray}
so that, as $r\rightarrow \infty$, $T\rightarrow v_0^4r/G$.

The virial theorem reads
\begin{equation}
{2T(r)+ \Omega(r)} = 4\pi r^3 \rho \langle v_r^2\rangle.
\end{equation}
As $r\rightarrow \infty$, the virial theorem does not take its familiar
form, but instead:
\begin{equation} 2T+\Omega \rightarrow
\begin{cases} 
{v_0^4r\over 2G}, & \mbox{if } \beta =0, \\
{v_0^4r\over G}, & \mbox{if } \beta =1/2.\end{cases}
\end{equation}
Just like the isothermal sphere, the models do not obey the virial
theorem unless the surface term is included.

\subsection{Projected and Line of Sight Quantities}

The projected surface density $\Sigma(s)$ is best written as
\begin{equation}
\Sigma(R) = {v_0^2\over 2 \pi G a (s^2-1)}(1 + (s^2-2)X(s)).
\end{equation}
Here, $s = R/a$, and, following Hernquist (1990), we have used the
notation
\begin{equation}
X(s) = \begin{cases}
{\displaystyle {1\over \sqrt{1-s^2}}\arcsech s}, & 0\le s\le 1,\nonumber\\
{\displaystyle {1\over \sqrt{s^2-1}}\arcsec s }, & s\ge 1.
\end{cases}
\end{equation}
Note that $X(1)=1$.

At the centre, $\Sigma(R)$ diverges logarithmically, whereas it falls
like $1/R$ at large radii, viz.,
\begin{eqnarray}
\Sigma_{\rm cusp}(R) &=& {v_0^2\over \pi G a} \log \left( {2a \over R}
\right),\nonumber \\
\Sigma_{\rm asym}(R) &=& {v_0^2\over 4 G } {1\over R}.
\end{eqnarray}

The line of sight or projected velocity dispersion of the isotropic
model ($\beta=0$) is
\begin{eqnarray}
\langle v_{\rm los}^2 \rangle &=& {2\over \Sigma(R)} \int_R^\infty \rho\langle
v_r^2\rangle {rdr\over \sqrt{R^2-r^2}}\nonumber\\
&=& {v_0^2\over 2}{(4s^2 + 2\pi s
  -3 -2\pi s^3) + s^2(4s^2-5)X(s) \over 1 + (s^2-2)X(s)}
\end{eqnarray}
which tends to $v_0/\sqrt{2}$ at large radii. At small radii, the
projected velocity dispersion tends to zero logarithmically
\begin{equation}
\langle v_{\rm los}^2 \rangle \rightarrow {3v_0^2\over 4} {1\over
 \log(2a/R)}
\end{equation}

Finally, the line of sight velocity dispersion of the radially
anisotropic ($\beta = 1/2$) model is
\begin{eqnarray}
\langle v_{\rm los}^2 \rangle &=& {2\over \Sigma(R)} \int_R^\infty \rho\langle
v_r^2\rangle \left(1- {R^2\over 2r^2}\right){rdr\over \sqrt{R^2-r^2}}\\
&=& {v_0^2\over 4}{(5 + 2\pi s^£ -4s^2-2\pi s) - (6-9s^2 + 4s^4)X(s) \over 1 + (s^2-2)X(s)}\nonumber
\end{eqnarray}
At both small and large radii, the velocity dispersion tends to the
value $\sqrt{3} v_0/2$. The intrinsic and line of sight velocity
dispersions are shown in Fig.~\ref{fig:vel}.

\begin{figure}
\includegraphics[width=3.25in]{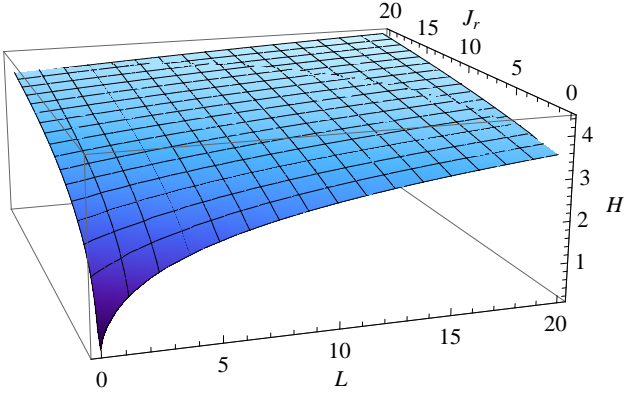}
\caption{A surface plot of $\mathcal{H}$ for the case 
$a=v_0=1$.}
\label{fig:Hsurf}
\end{figure}

\subsection{The Hamiltonian and the Actions}

Action-angles are an extremely useful set of coordinates in which to
visualize the form of a Hamiltonian. Since actions are integrals of
motion, they automatically satisfy $\dot{J_{i}}=0$, and the
canonically conjugate angle variables evolve in time in a trivial
fashion:
\begin{equation}
\theta_{i}(t)=\theta_{i}(0)+\Omega_{i}t,\qquad\qquad
\Omega_{i} = \frac{\partial H}{\partial J_{i}}.
\end{equation}
In a spherical potential, the radial action $J_{r}$ is:
\begin{equation}
J_{r}=\oint\sqrt{2\psi- 2E-L^{2}/r^{2}}\mathrm{d}r,
\label{eq:Jr}.
\end{equation}
The angular actions are related to the angular momentum components via
$J_\phi = L_z$ and $J_\theta = L - L_z$, where $L$ is given by
eq.~(\ref{eq:angmom}), whilst $L_{z}$ is the angular momentum about
the $z$-axis.  We then have a complete set of canonical coordinates
when the actions are complemented by their respective angles. In the
spherical case, the Hamiltonian can then be regarded as a function of
just two quantities: $H\equiv H(L,J_{r})$ (see Binney \& Tremaine
2008). However, due to the difficulty in evaluating the integral in
equation~(\ref{eq:Jr}), we cannot generally find $J_{r}(E,L)$ and
therefore $H(L,J_{r})$. In fact, Evans, Lynden-Bell \& de Zeeuw (1990)
showed that the most general potential for which the actions are
available as elementary functions is the isochrone.

For our model, it is therefore no surprise that $J_{r}(E,L)$ and
$L(E,J_{r})$ cannot be found in the general case. However, progress
can be made by using the approximations found in Williams, Evans \&
Bowden (2014) (hereafter WEB).  The method described in WEB
approximates $H(L,J_r)$ for scale-free spherical potentials by
calculating the energy of purely radial orbits
\begin{equation}
J_{r} = \dfrac{1}{2\pi}\oint\sqrt{2\psi(r)-2E(0,J_{r})}\mathrm{d}r,
\end{equation}
and of circular orbits
\begin{equation}
L(r) = rv_{\rm c}(r), \qquad\qquad E(L,0) = \psi(r)-\frac{1}{2}v_{\rm c}(r)^{2}.
\end{equation}
Each of these equations gives a solution with the same functional form, $-E(J_{i},J_{j}=0) = f(a_{i}J_{i})$, 
and WEB create an approximate Hamiltonian by analytically interpolating between these 
two limiting cases in a very simple way, so that
\begin{equation}
\mathcal{H}(L,J_{r}) = f(aL+bJ_{r}).
\label{eq:approx}
\end{equation}
Of course, the potential considered here is spherically symmetric but
not scale free. For example, the radial action of a general radial
orbit is given by
\begin{equation}
J_{r} = \frac{\sqrt{2}av_{0}}{\pi}\exp(-E/v_{0}^{2})\gamma(3/2,-E/v_{0}^{2}),
\label{eq:Jrgen}
\end{equation}
where $\gamma(s,x)$ is the lower incomplete gamma function. This
integral equation cannot be solved in the general case, so we look to
find specific solutions in the limiting cases. The obvious limits to
consider are orbits in the cusp of the model ($r\ll a$) and orbits in
the outer halo ($r \gg a$). Using these limiting cases, we can then
use analytical interpolation to find a function $\mathcal{H}$ that
well approximates the true Hamiltonian $H$ everywhere in action space.
   
\subsubsection{The Cusp}

In the cusp of the potential ($r\ll a$), we Taylor expand the
potential as
\begin{equation}
\psi(r) = -\dfrac{v_0^2 r}{a} + \mathcal{O}(r^2),
\label{eq:taylorcusp}
\end{equation} 
demonstrating that the potential is a scale-free power-law potential
in this regime. As a result, we can employ the results from WEB to
find an accurate analytical form for $H_{\mathrm{cusp}}(L,J_r)$. We
use their equation (12) with $\alpha=1$ to find
\begin{equation}
\mathcal{H}_{\mathrm{cusp}}(L,J_r) = \dfrac{3v_0^{4/3}}{2a^{2/3}}\bigg(L+\frac{\pi}{\sqrt{3}}J_r\bigg)^{2/3}.
\end{equation}
For a more accurate result, we can employ the first-order corrected
version of this formula to give
\begin{equation}
\mathcal{H}_{\mathrm{cusp}}(L,J_r) = \dfrac{3v_0^{4/3}}{2a^{2/3}}\bigg(L+\frac{\pi}{\sqrt{3}}J_r+\epsilon\sqrt{LJ_r}\bigg)^{2/3},
\end{equation}
where $\epsilon=-0.0412$. This approximation is valid for orbits that
spend the majority of their time in the region in the cusp regime.
Numerically, we find that this corresponds to
$|\boldsymbol{J}|=\sqrt{L^2+J_r^2}\lesssim0.06av_0$. Given this
expression for the Hamiltonian, we may then approximate the
distribution function in this region of phase space by substituting
$\mathcal{H}_{\mathrm{cusp}}$ into equation (\ref{eq:DFlim}):
\begin{equation}
\Fc(L,J_r)= \mathcal{N}L^{2\beta}\bigg(L+\frac{\pi}{\sqrt{3}}J_r+\epsilon\sqrt{LJ_r}\bigg)^{-5/3 + 2\beta},
\end{equation}
where $\mathcal{N}$ is the normalisation factor.

\subsubsection{The Far-Field}

If we instead consider the case when $r\gg a$, then the potential
takes the form
\begin{equation}
\psi(r) \simeq -v_0^2\log r/a,
\end{equation}
so this potential coincides with the singular isothermal sphere at
large radii. In this case, the approximate Hamiltonian is given by
\begin{equation}
\mathcal{H}_{\mathrm{asym}}(L,J_r) = v_0^2\log\bigg(\dfrac{\sqrt{e}L+\sqrt{2\pi}J_r}{av_0}\bigg).
\end{equation}
As in the cusp, we can also improve this approximation if need be. The
first-order corrected formula is
\begin{equation}
\mathcal{H}_{\mathrm{asym}}(L,J_r) = v_0^2\log\bigg(\dfrac{\sqrt{e}L+\sqrt{2\pi}J_r+\epsilon\sqrt{LJ_r}}{av_0}\bigg)
\end{equation}
with $\epsilon=-0.1009$. If we insert this expression into the
far-field limit of the distribution function in eqn (\ref{eq:DFlim}),
we find
\begin{equation}
\Fa(L,J_r) = \mathcal{M} L^{2\beta}\bigg(\sqrt{e}L+\sqrt{2\pi}J_r+\epsilon\sqrt{LJ_r}\bigg)^{2-2\beta},
\end{equation}
where $\mathcal{M}$ is again the normalisation factor. Hence, we have
shown that, in both the cusp and outer halo of this potential, the
distribution function as a function of the actions is well
approximated by a simple power law.

\subsubsection{Generalisation to all action space}
\begin{figure}
	\includegraphics[width=4in]{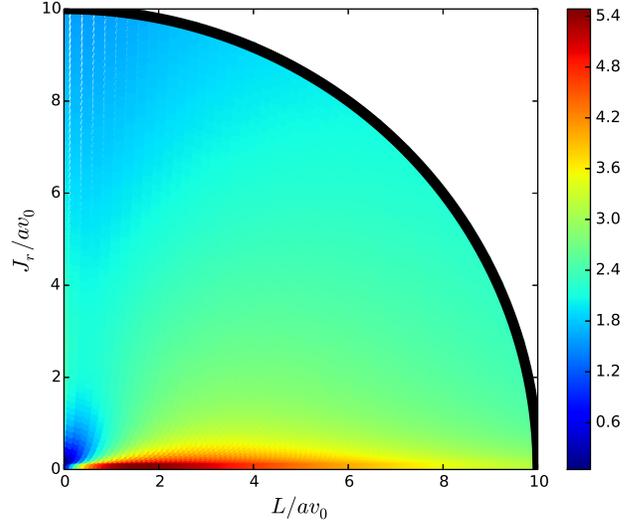}
	\caption{Distribution of absolute percentage error in $H$ in
          the region $0\leq|\boldsymbol{J}|\leq 10av_0$. One can see
          that the error peaks at $\sim 5$ percent close to the
          $L$--axis.}
	\label{fig:errtrans}
\end{figure}

Now that we know the functional forms of $\mathcal{H}_{\mathrm{cusp}}$
and $\mathcal{H}_{\mathrm{asym}}$, we are in a position to construct a
function $\Ha$ that approximates the full Hamiltonian $H$ everywhere
in action space. Consider the following function
\begin{equation}
\Ha(L,J_r) = \dfrac{3v_0^2}{2}\log\left(1 + \left(\dfrac{A(L)L + B(J_r)J_r}{av_0}\right)^{2/3}\right)
\label{eq:Htot}
\end{equation}
where
\begin{equation}
A(L) = \dfrac{\sqrt{e}L+av_0}{L+av_0}\quad;\quad B(J_r) = \dfrac{\sqrt{6\pi}J_r + \pi av_0}{\sqrt{3}(J_r+av_0)}.
\end{equation}
If we evaluate $\Ha(L,J_r)$ in the limits $|\boldsymbol{J}|<<av_0$ and $|\boldsymbol{J}|>>av_0$, we find:
\begin{equation}
\Ha(L,J_r)=\begin{cases} \dfrac{3v_0^{4/3}}{2a^{2/3}}\bigg(L+\dfrac{\pi}{\sqrt{3}}J_r\bigg)^{2/3}\quad |\boldsymbol{J}|<<av_0,  \\ \\
		  v_0^2\log\bigg(\dfrac{\sqrt{e}L+\sqrt{2\pi}J_r}{av_0}\bigg) \quad |\boldsymbol{J}|>>av_0.
    \end{cases}
\end{equation}
Hence, we find that $\Ha$ coincides with $\Hac$ and $\Haf$ in the
corresponding limits. The appearance of $\Ha$ can be seen in Figure
\ref{fig:Hsurf}. To test the accuracy of this approximation, we
generated a finely spaced grid in action space, with
$0\leq|\boldsymbol{J}|\leq 30av_0$. We find that the maximum absolute
error on our grid is 5.48 percent, and the mean error is 1.61
percent. Predictably, the errors are largest in the regime
$|\boldsymbol{J}|\sim av_0$, in which we transit between $\Hac$ and
$\Haf$, and lowest in the regimes where $\Hac$ and $\Haf$ apply. The
error distribution for the region $0\leq|\boldsymbol{J}|\leq 10av_0$
is given in Fig.~\ref{fig:errtrans}, where one can clearly see that
the error is worst close the the $L$--axis. This suggests that a
correction term in $L$ might be prudent in eq.~(\ref{eq:Htot}).


\begin{figure}
    \centering
    \includegraphics[width=3.25in]{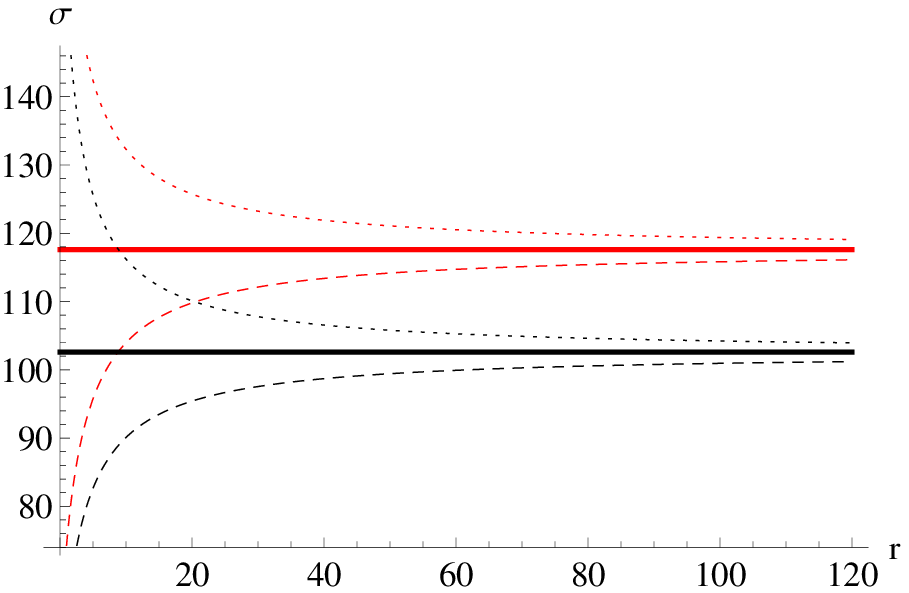}
    \includegraphics[width=3.25in]{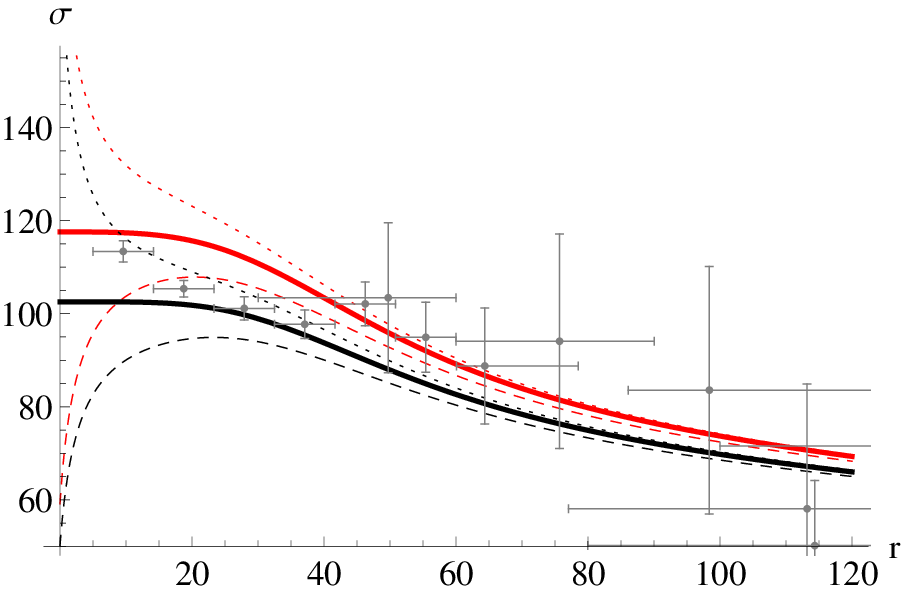}
    \includegraphics[width=3.25in]{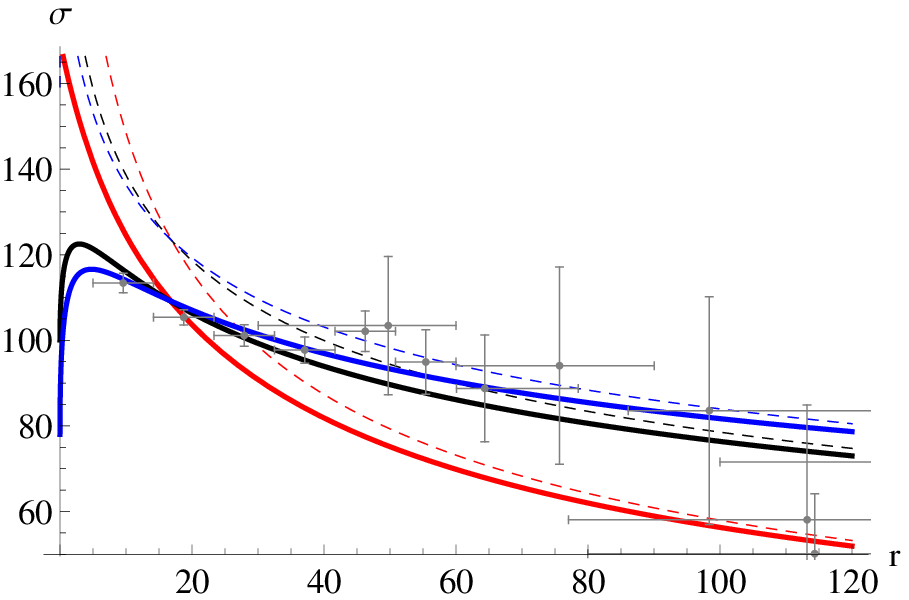}
    \caption{Radial velocity dispersion (in kms${-1}$) as a function
      of spherical polar radius $r$ (in kpc). Upper panel: When the
      rotation curve is flat, the radial velocity dispersion of an
      asymptotically power-law population is constant, as shown for
      $n= 3.5$ (red) and $n=4.6$ (black). If the lengthscales of the
      dark matter is greater than (less than) that of the luminous
      matter, then the constant value is approached from below
      (above), as shown by the dashed (dotted) lines. Middle panel:
      The power-law densities are now modulated by a taper function
      that causes exponential decline beyond the taper radius $\rt$,
      as in eq.~(\ref{eq:taper}). The effect of truncation of the
      density of the stellar population causes the isotropic velocity
      dispersion to decline. The colour coding is the same as in the
      upper panel, and the truncation parameters are $\rt =40$ kpc and
      $p=10$.  In this and the following panel, the datapoints of
      Deason et al. (2012) are overplotted in grey.  Lower panel: The
      populations have Einasto profiles with an effective radius $\re
      =20$ kpc and isotropic velocity distributions. Red, black and
      blue curves refer to Einasto indices of 1, 1.7 and 2
      respectively. Full (dashed) curves denote the case when the dark
      matter lengthscale $a$ is 1 (7.5) kpc. An Einasto model with
      $m=2$ and $\re =20$ kpc gives a good description of Deason et
      al's data out to $\sim 110$ kpc.}
    \label{fig:tracers}
\end{figure}

\begin{figure}
    \centering
    \includegraphics[width=3.25in]{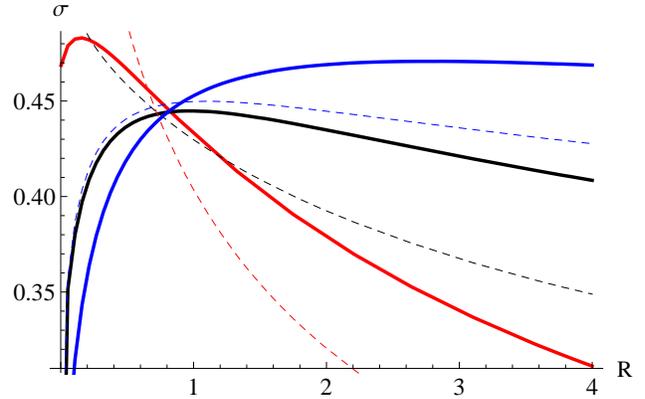}
\caption{Line of sight velocity dispersion (in units of $v_0$) versus
  projected radius (in units of effective radius $R_{\rm e}$) for
  Sersic tracers. The isotropic cases are shown as thick lines, the
  radial anisotorpic cases as dotted. The models vary from exponential
  or $m=1$ (red), $m=2$ (black) to de Vaucouleurs or $m=4$ (blue).}
\label{fig:sersicvels}
\end{figure}

\section{Tracer Populations}

So far, we have given properties of the self-consistent model. From
the point of view of comparison with observational data, it is
interesting to look at the properties of tracer populations. These
might represent the stellar halos of disc galaxies or even entire
elliptical galaxies embedded in a dark halo.  In what follows, the
total potential (dark matter and baryons) is always described by
eq.~(\ref{eq:newlog}), whilst the kinematic properties of the baryons
are deduced from their light profiles using the Jeans equations.

\subsection{Stellar Haloes}

First, let us assume the stars follow the density-law
\begin{equation}
\rho_\star = {\rho_0 b^n\over (b +r)^n}.
\end{equation}
Here, $b$ is the lengthscale of the luminous matter, as compared to
$a$ which is the lengthscale in the potential, and hence of the dark
matter. Power-laws are often used to describe populations in stellar
haloes. Traditionally, in our own Galaxy, the density of Population II
stars has often been described as a power-law with $n\approx 3.5$ (e.g.,
Freeman 1987). However, recent studies tend to find a faster fall-off
(e.g., Watkins et al. 2009, Sesar et al. 2011).  For example, using
blue horizontal branch and blue straggler stars extracted from the
Sloan Digital Sky Survey data, Deason et al. (2011) showed that the
stellar halo is well fitted by a power-law with $n\approx 4.6$ for
Galactocentric radii beyond $\approx 27$ kpc. Re-visiting the problem
with Sloan Data Releases 9, Deason et al. (2014) argue that an even
steeper power-law slope ($n \sim 6-10$) is favoured at larger radii ($r
\sim 50-100$ kpc).

When $a=b$, the isotropic ($\beta=0$) distribution function of the
stars is an isothermal and therefore the velocity distribution is a
Gaussian with dispersion $v_0/\sqrt{n}$.  This property of isothermals
was already known to Smart (1938).  More generally, for arbitrary
anisotropy $\beta$ and arbitrary lengthscales for the dark and
luminous matter $a$ and $b$, we can derive the asymptotic result
\begin{equation}
\langle v_r^2 \rangle = {v_0\over \sqrt{n-\beta}}\left[1  + {n(b-a) +
  a\beta \over 2(n+1-\beta)} {1\over r}\right].
\label{eq:asymp}
\end{equation}
{\it The radial velocity dispersion of an asymptotically power-law
  population in a galaxy with a flat rotation-curve always tends to
  a constant value}.  This finding is unaffected by velocity
anisotropy $\beta$, which simply changes the constant asymptotic value
to $v_0/\sqrt{n-\beta}$. For the isotropic case, the radial velocity
dispersion reaches its asymptotic value from above (below) according
to whether the lengthscale of the dark matter is less than (greater
than) the lengthscale of the luminous matter. This can be deduced from
eq~(\ref{eq:asymp}) and is illustrated in the upper panel of
Fig.~\ref{fig:tracers}.

However, this is very different to the behaviour of the radial
velocity dispersion of populations in our own Galaxy's halo. Deason et
al. (2012) show that the radial velocity dispersion at 100 kpc has
fallen to at least 40 per cent of its central value (see their Figure
9). If the rotation curve is flat, then it immediately follows that
the stellar halo cannot have any power-law decline, but must fall off
much more abruptly. This is because there is no sign in the data of a
velocity dispersion tending to a constant value, at least within 100
kpc.

To illustrate this, let us consider truncated power-laws of the form
\begin{equation}
\rho_\star = {\rho_0 b^n\over (b +r)^n} \Biggl[ 1-
  \tanh^p(r/\rt)\Biggr].
\label{eq:taper}
\end{equation}
Here, $\rt$ is a truncation radius. For $r\ll \rt$, the tracer density
has a power-law decline, but for $r\gg \rt$ the power-law is modulated
by a faster exponential decline with scalelength $\rt/p$. The higher
the value of $p$, the sharper the truncation.  The middle panel of
Fig.~\ref{fig:tracers} shows the radial velocity dispersion, inferred
from numerical integrations. To obtain the decline reported by Deason
et al. (2012), the stellar halo must be abruptly truncated ($p=10$)
beyond $\rt =40$ kpc.
 
Another possibility is that the stellar halo is described by an
Einasto profile
\begin{equation}
\rho_\star(r) = \rho_0\exp\Biggl[ -d_m[(r/\re)^{1/m} -1]\Biggr],
\end{equation}
where $\re$ is the effective radius and the constant $d_m$ is
well-approximated by (e.g., Merritt et al. 2006, Mamon \& Lokas 2005)
\begin{equation}
d_m = 3m - 1/3 - 0.0079/m.
\end{equation}
Deason et al. (2012) considered this possibility and found that $\re
=20$ kpc and $m = 1.7$ provided the best fit to the SDSS starcount
data.  As shown in the final panel of Fig.~\ref{fig:tracers}, the
radial velocity dispersons of Einasto profiles can provide good
matches to the kinematic data provided $m\approx 1.7-2$. Slightly
higher Einasto indices than 1.7 provide a somewhat better fit (blue
curve), as the dispersion decreases less rapidly at large radii. This
conclusion is supported by the most recent analysis of A-type stars
from SDSS (Deason et al. 2014). In fact, these authors do not fit an
Einasto profile to the stellar density data, but rather a broken
power-law model with index $n \approx 5$ favoured at 50 kpc and still
steeper indices at larger radii. However, Deason (2014, private
communication) finds that the steepening of the density law found with
the deepest SDSS data can also be extremely well-fit by an Einasto
profile with $m\approx 1.7-2$.

It is striking that the power-law and Einasto profiles both give very
good matches to the SDSS starcount data, but provide rather different
predictions for the velocity dispersions. A combination of photometric
and kinematic data is therefore a much more discriminating diagnostic
than starcount data alone.

\subsection{Elliptical Galaxies}

Finally, we discuss the line of sight velocity dispersion of tracer
populations, which has applications to populations (e.g., globular
clusters or planetary nebulae) in the outer parts of elliptical
galaxies.

We start by assuming that the light profile is a power-law
\begin{equation}
I_\star(R) = I_0 {b^{n-1} \over (b+R)^{n-1}},
\end{equation}
where $R$ is again the projected radius. This is inspired by analogy
with eq.~(\ref{eq:asymp}). The three dimensional density $\rho_\star$
falls off asymptotically like $r^{-n}$. We can show by solving the
spherical Jeans equation that the line of sight velocity dispersion
tends to a constant value at large projected radii
\begin{equation}
\sigma_{\rm p}^{2} = v_0^2 {n+ \beta(1-n) \over n(n-2\beta)}
\end{equation}
{\it In other words, the line of sight velocity dispersion of an
  asymptotically power-law population in a galaxy with a flat
  rotation-curve always tends to a constant value}. Again, this result
is unaffected by velocity anisotropy $\beta$, which merely changes the
constant value. This is the projected analogue of our earlier result.

Now, we will assume the light profile is a Sersic (1968) Law
\begin{equation}
I_\star(R) = I_0\exp\Biggl[ -d_m[(R/\re)^{1/m} -1]\Biggr].
\end{equation}
Of course, this gives an exponential if $m=1$ and a de Vaucouleurs
profile if $m=4$. We shall use the helpful approximation developed by
Prugniel \& Simien (1997), namely
\begin{equation}
d_m = 2m - 1/3 + 0.009876/m.
\end{equation}
Agnello, Evans \& Romanowsky (2014) showed how to calculate the line
of sight velocity dispersion from the surface brightness without the
rigmarole of deprojection, solution of the Jeans equations and
subsequent reprojection. Using their equation (10), we have
\begin{eqnarray}
\nonumber I_\star \sigma_{\rm p}^{2}(R)=\frac{2G}{\pi}\int_{R}^{\infty}s
I_\star(s)
\int_{R}^{s}\frac{\partial_{r}\left(M(r)\sqrt{r^{2}-R^{2}}/r^{3}\right)}{\sqrt{s^{2}-r^{2}}}\mathrm{d}r\mathrm{d}s\ \\ +
\frac{2G}{\pi} \int_{R}^{\infty}sI_\star(s)\int_{R}^{s}\frac{\partial_{r}
  \left(M(r)k_{\beta}(R,r)/r^{3}\right)}{\sqrt{s^{2}-r^{2}}}\mathrm{d}r\mathrm{d}s\ .
\label{eq:projsigma}
\end{eqnarray}
For our model, the enclosed mass is
\begin{equation}
M(r) = {v_0^2 \over G} {r^2\over a+r}.
\end{equation}
When the model is isotropic ($\beta =0$), then $k_\beta =0$ and only
the first term in eq~(\ref{eq:projsigma}) remains. For radial
anisotropy ($\beta = 1/2$), we have:
\begin{equation}
k_\beta(R,r) =r\log\left( {r+\sqrt{r^2-R^2}\over R}\right) -
\frac{3}{2}\sqrt{r^2-R^2}
\end{equation}
Fig.~\ref{fig:sersicvels} shows the line of sight velocity dispersion
as a function of radius for Sersic models with $m=1$ (exponetial law),
$m=2$ and $m=4$ (de Vaucoulers law) for the isotropic and radially
anisotropic cases. The profiles always fall at large radii, with the
radially anisotopic models ($\beta = 1/2$) declining more strngly than
the isotropic. The behaviour at the centre is may be declining or
increasing, though this will be smeared out by the finite aperture
size for real data.

\section{Conclusions}

We have presented a new and very simple halo model. It has an flat
rotation curve at large radii, but the matter density is cusped like
$1/r$ at small radii. We have found simple isotropic and radially
anisotropic phase space distribution functions (DFs), whilst the
second moments and projected kinematic quantities are analytic.  The
new model is closely related to Hernquist's (1990) model, from which
it can be derived by the same procedure used recently by Evans \&
Bowden (2014) to find the missing model in the Miyamoto \& Nagai
(1975) sequence. The model also has properties in common with the
famous NFW profile, which is the endpoint of collisionless dark matter
simulations in hierarchically merging cosmologies. However, it is much
simpler than the NFW profile, whose awkward potential means that some
properties, such as the DFs or the actions, can be calculated only
numerically (e.g., Lokas \& Mamon 2001).

It is helpful to compare our new model with two other widely used
representations of haloes.  The first is the spherical limit of
Binney's logarithmic potential (Binney 1981, Evans 1993, Binney \&
Tremaine 2008)
\begin{eqnarray}
\psi &=& -{v_0^2\over 2} \log (r^2+a^2)/a^2,\nonumber\\
\rho &=& {v_0^2 \over 4\pi G}{r^2 + 3a^2\over (r^2+a^2)^2}.
\label{eq:binney}
\end{eqnarray}
Note the main difference is that our model (\ref{eq:newlog}) has a
$1/r$ density cusp, whereas (\ref{eq:binney}) is cored at the
centre. The properties at large radii of both models are similar, as
they both tend towards isothermal. The second is Jaffe's (1983) model,
which has potential-density pair:
\begin{eqnarray}
\psi &=& v_0^2\log \left[ {a+r\over r}\right] = v_0^2\log (a+r) -
v_0^2\log r \nonumber \\
\rho &=& {v_0^2 \over 4\pi G}{a^2\over r^2(a+r)^2}.§
\end{eqnarray}
So, the model introduced in this paper is therefore the difference of the
Jaffe model and the isothermal sphere!

We note that there is still scope for further exploration in this
area. Binney's spherical logarithmic potential, the model in this
paper, the singular isothermal sphere and the Jaffe model form part of
a larger family with
\begin{equation}
\psi = -{v_0^2\over p} \log (r^p+a^p)/a^p,
\end{equation}
when $p = 2,1,0$ and $-1$ respectively. The properties of this large
class of halo models -- {\it the doubloon family} -- we investigate in
a related publication.

As an application, we have considered tracer populations in the outer
parts of the model. With power-law or Einasto density profiles, these
may represent population II stars or globular clusters in the stellar
halo of a spiral galaxy like our own.≈ß With Sersic surface brightness
profiles, they may represent stellar populations in elliptical
galaxies. We show that -- for a galaxy with a flat rotation curve --
density laws that tend to power-laws at large radii always have
asymptotically constant radial velocity dispersion, irrespective of
the anisotropy or the lengthscales characteristic of the luminous and
dark matter. We use this result to demonstrate that the stellar halo
of our own Galaxy must fall off more quickly than a
power-law. Truncated power-laws and Einasto profiles give a much
better representation of the observed decline of the velocity
dispersion. {\it In particular, both the starcount and kinematic data of
the Milky Way stellar halo are well-represented by an Einasto profile
with index $m\approx 2$ and effective radius $\approx 20$ kpc, if the
dark halo has a flat rotation curve.}

\section*{Acknowledgments}
We thank Adam Bowden and Alis Deason for interesting conversations,
and the anonynmous referee for a valuable report. AW is supported by
the Science and Technology Facilities Council (STFC) of the United
Kingdom.

\bibliography{miynag}
\bibliographystyle{mn2e}

\appendix

\section{Auxiliary Formulae}

Here, we list the general formulae for the constants in the
distribution function (\ref{eq:dffull}):
\begin{eqnarray}
A &=& { (2-2\beta)^{3/2-\beta}\over 2^{7/2 -\beta}\pi^{5/2}\Gamma(1-\beta)
Ga (av_0)^{1-2\beta}}\nonumber\\
C_j &=&  {\displaystyle (-1)^j(1+ \frac{j}{2-2\beta})^{3/2-\beta}\left[
  {2\beta-1\choose j} - {2\beta-1,\choose j-1}\right]}
\end{eqnarray}
The cases $\beta=0$ and $\beta = 1/2$ are discussed in the body of the
paper.

\end{document}